\documentclass[twocolumn]{revtex4}

\usepackage{graphicx}
\usepackage{dcolumn}
\usepackage{bm}
\usepackage{amsmath}
\usepackage{amsfonts}
\begin{document}

\title{Quantum model of microcavity intersubband electroluminescent devices}
\author{Simone \surname{De Liberato}$^{1,2}$}
\author{Cristiano Ciuti$^{1}$}
\email{cristiano.ciuti@univ-paris-diderot.fr}
\affiliation{$^1$Laboratoire Mat\'eriaux et Ph\'enom\`enes
Quantiques, Universit\'e Paris Diderot-Paris 7 and CNRS, UMR 7162, \\
B\^atiment Condorcet, 75205 Paris Cedex 13, France}
\affiliation{$^2$Laboratoire Pierre Aigrain, \'Ecole Normale
Sup\'erieure, 24 rue Lhomond, 75005 Paris, France}

\begin{abstract}
We present a quantum theoretical analysis of the
electroluminescence from an intersubband transition of a quantum
well structure embedded in a planar microcavity. By using a
cluster factorization method, we have derived a closed set of
dynamical equations for the quantum well carrier and cavity photon
occupation numbers, the correlation between the cavity field and
the intersubband polarization, as well as
polarization-polarization contributions. In order to model the
electrical excitation, we have considered electron population
tunneling from an injector and into an extractor contact. The
tunneling rates have been obtained by considering the bare
electronic states in the quantum well and the limit of validity of
this approximation (broad-band injection) are discussed in detail.
We apply the present quantum model to provide a comprehensive
description of the electronic transport and optical properties of
an intersubband microcavity light emitting diode, accounting for
non-radiative carrier relaxation and Pauli blocking. We study the
enhancement of the electroluminescence quantum efficiency passing
from the weak to the strong polariton coupling regime and compare 
it with the free-space case.
\end{abstract}
\maketitle

In the last two decades, the fundamental research on the physics
of intersubband transitions in semiconductor quantum wells has
enjoyed a considerable success and also led to novel applications
in quantum optoelectronics\cite{ISB-Book}. Recently, reflectivity
experiments \cite{Dini_PRL,Aji_APL,Aji_2006} have demonstrated
that by embedding a doped quantum well structure in a planar
microcavity, it is possible to achieve the strong coupling regime
between an intersubband transition and a cavity photon mode, provided that a dense enough
two-dimensional electron gas populates the fundamental quantum
well subband. The interaction between a bright intersubband
excitation and a cavity photon is quantified by the so-called
vacuum Rabi frequency. The
strong coupling regime occurs when the vacuum Rabi frequency exceeds the electronic and photonic losses. In such a regime, the normal modes of the system are cavity polaritons, half-photon half-intersubband excitations. In this kind of system, it is even possible to reach an unconventional ultra-strong coupling regime, i.e. a vacuum Rabi frequency comparable to the intersubband transition
frequency\cite{Ciuti_vacuum,Ciuti_PRA,Simone_PRL} .

The interplay between judiciously quantum engineered intersubband
transitions and vertical electron transport is the essence of the
so-called quantum cascade electroluminescent devices and lasers,
which are unipolar optoelectronic sources emitting in the mid and
far infrared portion of the electromagnetic spectrum
\cite{Faist,Terahertz,Raffaele}. A new kind of
microcavity-embedded quantum cascade devices in the strong
coupling regime was proposed in Ref. \onlinecite{SST}. The first
experimental demonstrations of a microcavity quantum cascade
photovoltaic detector\cite{Luca_APL} and of an electroluminescent
device in the strong coupling regime have been recently
reported\cite{Luca-unpublished}.

This promising research topic is in its very infancy and many
interesting theoretical questions need to be addressed. In Ref.
\onlinecite{Ciuti_PRA}, intersubband polariton electroluminescence
has been analytically treated within a simplified Hamiltonian
model based on the following assumptions: (i) only the bright
intersubband excitations have been taken into account, while dark
excitations have been neglected; (ii) only the low excitation
regime has been considered, in which the bright intersubband
excitations have been approximated as bosons; (iii) the electronic
coupling to the intersubband polarization field has been modeled
through a phenomenological reservoir of bosonic excitations. In
this work, we will attempt to treat the same problem starting
directly from the fermionic Hamiltonian for the quantum well
carriers. This approach can give us useful insight to understand
the physics obtained by relaxing the assumptions used in Ref.
\onlinecite{Ciuti_PRA} and to grasp which intrinsic factors
ultimately determine the quantum efficiency of these strong
coupling emitters.  On one hand, the large values of the vacuum
Rabi frequency could induce a very fast and efficient emission of
photons. On the other hand, the large density of dark intersubband
excitations created by the injection current and the Pauli
blocking in the densely populated fundamental subband could
suppress such enhancement.

We would like to point out that from a theoretical point of view, a
description of the considered system in terms of the fermionic
carrier operators makes the system Hilbert space much larger than
within a bosonic model. In this paper, we have followed an
approach based on a truncation of the infinite hierarchy of
dynamical equations for the operator expectation values, allowing
us to describe many relevant aspects of the intersubband
microcavity electroluminescence. However, some of the
non-perturbative features obtained analytically within a bosonic
model\cite{Ciuti_PRA} can not be accounted for within the present
treatment. Different fermionic approaches based on exact
diagonalization methods \cite{Deliberato_exact} are eventually
necessary for further refinements.

In this paper, we present a quantum model of the
spontaneous photon emission from an electrically-excited
intersubband transition of a quantum well structure embedded in a
planar microcavity mode. Here, we will consider the case of an
incoherent electron transport, where the quantum well electron
populations in the two subbands have a tunneling coupling to an
electronic injector and to an extractor. The tunneling rates have
been obtained by considering the bare electron states inside the
quantum well. The domain of validity of this approximation will be
discussed in detail. The present theoretical model is applied to
describe the incoherent electron transport and electroluminescence
of an intersubband microcavity light emitting diode in the strong
coupling regime. The paper is structured as follows. In Sec.
\ref{Ham}, we describe the system and introduce the second
quantization Hamiltonian describing electrons in the two
conduction subbands and photons in the fundamental microcavity
mode. In Sec. \ref{onetime}, we present a closed set of dynamical
equations for the one-time expectation values of operator
products, describing photon and carrier populations as well as
intersubband polarization-polarization and polarization-field
correlations. These equations have been obtained through a cluster
expansion, whose details are reported in Appendix \ref{app}. In
Sec. \ref{steady}, we discuss the steady-state regime obtained
under constant electrical excitation. The corresponding set of
algebraic equations for the steady-state expectation values are
reported in Appendix \ref{algebraic}. In Sec. \ref{spectra}, the
electroluminescence spectra are analytically calculated as a
function of the populations and the appearance of intersubband
cavity polaritonic resonances in the emission spectra is shown.
Numerical applications of the theory are presented in Sec.
\ref{numerical}, using a specific configuration for the injection
and extraction electronic reservoirs. The results predict the
current-voltage characteristics, emission spectra and quantum
efficiency using different (controllable) parameters for the
considered microcavity system. The results are critically
discussed with respect to the approximations of the model.
Finally, conclusions and future perspectives are drawn in Sec.
\ref{conclusions}.

\section{Description of the system and quantum Hamiltonian}
\label{Ham} The system under study is described by the
following second quantization Hamiltonian
\begin{widetext}
\begin{eqnarray}
H&=&\sum_{\mathbf{k},\sigma}\hbar \omega_{1}(\mathbf{k})c_{1,\sigma,\mathbf{k}}^{\dagger}c_{1,\sigma,\mathbf{k}}+
\sum_{\mathbf{k},\sigma}\hbar\omega_{2}(\mathbf{k})c_{2,\sigma,\mathbf{k}}^{\dagger}c_{2,\sigma,\mathbf{k}}+
\sum_{\mathbf{k}}\hbar\omega_{c}(\mathbf{k})a_{\mathbf{k}}^{\dagger}a_{\mathbf{k}}\nonumber \\
&&+\sum_{\mathbf{k,q},\sigma}\hbar\chi(\mathbf{q})a_{\mathbf{q}}c_{1,\sigma,\mathbf{k}}c_{2,\sigma,\mathbf{k+q}}^{\dagger}
+\sum_{\mathbf{k,q},\sigma}\hbar\chi^{*}(\mathbf{q})a_{\mathbf{q}}^{\dagger}c_{2,\sigma,\mathbf{k+q}}c_{1,\sigma,\mathbf{k}}^{\dagger}
+ H_{other}.
\end{eqnarray}
\end{widetext}
The energy dispersions of the two quantum well conduction subbands
are $\hbar \omega_{1}(\mathbf{k}) =  \frac{\hbar^2
k^2}{2m^\star}$ and
 $\hbar \omega_{2}(\mathbf{k}) = E_{12} + \frac{\hbar^2 k^2}{2m^\star}$, being  $\mathbf{k}$ the electron in-plane wavevector and $m^\star$ the effective
mass (non-parabolicity is here neglected).
 The corresponding electron creation fermionic operators are $c_{1,\sigma,\mathbf{k}}^{\dagger}$ and $c_{2,\sigma,\mathbf{k}}^{\dagger}$, where $\sigma$ is the electron spin.
 $\omega_{c}(\mathbf{q}) = \frac{c}{\sqrt{\epsilon_r}} \sqrt{q_z^2+q^2}$ is the bare frequency dispersion of a cavity photonic branch as a function of the in-plane
 wavevector $\mathbf{q}$, where $c$ is the light speed, $\epsilon_r$ is the cavity spacer dielectric constant and $q_z$ is the quantized photon wavevector along the normal
direction. $a_{\mathbf{q}}^{\dagger}$ is the corresponding photon
creation operator, obeying bosonic commutation rules. Due to the
well-known polarization selection rules of intersubband
transitions, we omit the photon polarization, which is assumed to
be Transverse Magnetic (TM). For simplicity, we consider only a
photonic branch, which is quasi-resonant with the intersubband
transition, while other cavity photon modes are supposed to be
off-resonance and can be therefore neglected in first
approximation.
 The interaction between the cavity photon field and the
two electronic subbands is quantified by the coupling constant
\begin{equation}
\label{chi}
\chi(\mathbf{q}) = \sqrt{\frac{\omega_{12} ^2d_{12}^2}{\hbar \epsilon_0 \epsilon_r L_{cav} S \omega_{c}(\mathbf{q})} \frac{q^2}{(\pi/L_{cav})^2 + q^2 } },
\end{equation}

where $d_{12}$ is the intersubband transition dipole along the
quantum well growth direction, $\omega_{12} = E_{12}/\hbar$ is the frequency of the intersubband transition,
$\epsilon_0$ the vacuum permittivity, $L_{cav}$ is the effective
cavity length and $S$ is the sample area. For simplicity, we
have considered a $\lambda/2$-cavity, with $q_z =\pi/L_{cav}$
being the quantized vector along the growth direction. The
geometrical factor $\frac{q^2}{(\pi/L_{cav})^2 + q^2 }$ originates
from the TM-polarization nature of the transition. Moreover, in Eq.
(\ref{chi}) we have assumed that the active quantum well is located at the antinode of
the cavity mode field, providing maximum coupling. Note that here
we have neglected the anti-resonant terms of the light-matter
interaction and therefore we can describe the strong coupling
regime for the electrically excited system, but not the
ultrastrong coupling limit, as instead done in
Refs. \onlinecite{Ciuti_vacuum,Ciuti_PRA,Simone_PRL}. The
Hamiltonian term $H_{other}$ is meant to include all the other
interactions: (i) electron-phonon interaction; (ii)
electron-electron interaction; (iii) electron tunneling coupling
to the injection and extraction reservoir; (iv) coupling between
the cavity photon field and the extracavity field.

\section{Closed set of dynamical equations for the one-time expectation values}
\label{onetime} It is known that due to the cubic light-matter
coupling term in the Hamiltonian (the product of two fermion
operators and one boson operator) it is not possible to write down
an exact closed set of equations for the evolution of operators,
being the Heisenberg equation of motion for each product of $N$
operators coupled at least with one product of $N+1$ operators. In
other words, the equations of motion of the different observables
of the system form an infinite hierarchy. One approximation method
that has been used in order to solve this kind of systems is the
so-called cluster expansion scheme\cite{F1, F2, KK1}. It
is based on a systematic development of expectation values of
operator products in terms of correlation functions.

In order to obtain a consistent truncation scheme, a pair of fermionic
operators has to be considered of the same order as a single
bosonic operator.  In this work, we have truncated the hierarchy
at the level of the product of two excitation operators (i.e., the
product of four fermion operators). The details of the
factorization are in Appendix \ref{app}. The expectation values entering the
present cluster factorization are the electronic and photonic populations,
the correlation between the cavity photon field and the intersubband polarization, as well as polarization-polarization correlations. The electron occupation numbers in the
two quantum well conduction subbands are $n_{1,\mathbf{k}} =
<c_{1,\sigma,\mathbf{k}}^{\dagger}c_{1,\sigma,\mathbf{k}}>$ and
$n_{2,\mathbf{k}} = <c_{2,\sigma,\mathbf{k}}^{\dagger}c_{2,\sigma,\mathbf{k}}>$.  Note that, since in the absence of a magnetic field all quantities are spin-independent, we omit the spin-index in the notation of the averaged quantities. The cavity photon number is $n_{a,\mathbf{q}} = <a_{\mathbf{q}}^{\dagger}a_{\mathbf{q}}>$. The correlation between the cavity photon field and the
intersubband electronic polarization is represented by the quantity
\begin{equation}
Y(\mathbf{q,k}) =
<a_{\mathbf{q}}^{\dagger}c_{1,\sigma,\mathbf{k}}^{\dagger}c_{2,\sigma,\mathbf{k+q}}>.
\end{equation}
Finally, the polarization-polarization correlation function is given by
\begin{equation}
X(\mathbf{q+k', k', k}) =
\sum_{\sigma}<c_{2,\sigma,\mathbf{q+k'}}^{\dagger}c_{1,\sigma,\mathbf{k'}}c_{1,\sigma',\mathbf{k}}^{\dagger}c_{2,\sigma',\mathbf{k+q}}>.
\end{equation}
Note that in the spontaneous photon emission regime,
$Y(\mathbf{q,k})$ can not be factorized: in fact, spontaneous
emission is incoherent and $< a_{\mathbf{q}}> = 0$,
$<c_{1,\sigma,\mathbf{k}}^{\dagger}c_{2,\sigma,\mathbf{k+q}}> =
0$, meaning that the cavity field and the intersubband
polarization have no definite phase. Loss of coherence due to
dephasing processes and photonic losses is phenomenologically
quantified by the damping rate $\Gamma_Y$. Unlike
$Y(\mathbf{q,k})$, $X(\mathbf{k'+q,k',k})$ can be factorized in
products of non-zero lower-order expectation values of operators.
In fact, we have $X(\mathbf{k'+q,k',k}) =  2
n_{2,\mathbf{k+q}}(1-n_{1,\mathbf{k}}) \delta_{\mathbf{k,k'}} +
\delta X(\mathbf{k'+q,k',k})$. The first contribution is an
uncorrelated plasma term, while $\delta X(\mathbf{k'+q,k',k})$
describes the higher-order correlation, which can be destroyed by
dephasing processes quantified by the damping rate $\Gamma_X$.

The terms in $H_{other}$, namely the phonon scattering,
electron-electron interaction, the coupling to the contact
reservoirs and the coupling to the external electromagnetic field
will be treated in an effective way. The carrier non-radiative
relaxation (due to phonon-electron and electron-electron
scattering) is modeled in terms of a simple phenomenological
relaxation time $\tau_{\mathbf{k}}$. Note that the role of
Coulomb electron-electron interaction on intersubband transitions
has been studied, e.g., in Ref. \onlinecite{Coulomb}. In the case
of subbands with parallel parabolic dispersion (e.g., same
effective mass), Coulomb interaction produces a moderate
renormalization of the intersubband transition frequency
$\omega_{12}$ and of its oscillator strength, which will not be
accounted explicitly in the present work.

Let $n^0_{1,\mathbf{k}}$ and $n^0_{2,\mathbf{k}}$ be the
self-consistent local equilibrium occupation numbers. They are
given by Fermi-Dirac distributions:
\begin{eqnarray}
n^0_{1,\mathbf{k}}&=&\frac{1}{e^{\beta(\hbar \omega_1(\mathbf{k})-\epsilon_F)}+1}\nonumber,\\
n^0_{2,\mathbf{k}}&=&\frac{1}{e^{\beta(\hbar \omega_2(\mathbf{k})-\epsilon_F)}+1},
\end{eqnarray}
where $\beta = 1/(KT)$ is the Boltzmann thermal factor, and $\epsilon_F$ is the quantum well self-consistent Fermi level, such that:

\begin{equation}
 \sum_{\mathbf{k}} n_{1,\mathbf{k}} + n_{2,\mathbf{k}}
=\frac{Sm^*}{2\pi\hbar^2}\int_0^{\infty}d\epsilon
\frac{1}{e^{\beta(\epsilon-\epsilon_F)}+1}+\frac{1}{e^{\beta(\epsilon+E_{12}-\epsilon_F)}+1}.
\end{equation}

The two subbands are coupled to two electronic reservoirs, named
respectively left and right contacts. We will call
$\Gamma^{in}_{p,j,\mathbf{k}}$ the electronic tunneling rate into
the $\mathbf{k}$-mode of the subband $j=1,2$ from the reservoir
$p={\rm left,right}$. Analogously $\Gamma^{out}_{p,j,\mathbf{k}}$
is defined as the electronic tunneling rate from the $\mathbf{k}$-mode of the subband $j$ into the reservoir $p$.  The total
in-tunneling and out-tunneling rates are
$\Gamma^{in}_{j,\mathbf{k}}=\Gamma^{in}_{{\rm
left},j,\mathbf{k}}+\Gamma^{in}_{{\rm right},j,\mathbf{k}}$ and
$\Gamma^{out}_{j,\mathbf{k}}=\Gamma^{out}_{{\rm
left},j,\mathbf{k}}+\Gamma^{out}_{{\rm right},j,\mathbf{k}}$.

The resulting closed system of equations for the one-time
expectation values reads:
\begin{widetext}
\begin{eqnarray}
\label{full_system}
\frac{d}{dt}n_{a,\mathbf{q}} &=& -2\gamma~n_{a,\mathbf{q}}+2i\sum_{\mathbf{k}}
\chi^{*}(\mathbf{q})Y(\mathbf{q,k})+c.c.\nonumber \\
\frac{d}{dt} n_{1,\mathbf{k}}&=&-\frac{n_{1,\mathbf{k}}-n^0_{1,\mathbf{k}}}{\tau_{\mathbf{k}}} -\Gamma^{out}_{1,\mathbf{k}} n_{1,\mathbf{k}}+\Gamma^{in}_{1,\mathbf{k}} (1-n_{1,\mathbf{k}})+i\sum_{\mathbf{q}}
\chi^{*}(\mathbf{q})Y(\mathbf{q,k})+c.c.\nonumber\\
\frac{d}{dt} n_{2,\mathbf{k}}&=&-\frac{n_{2,\mathbf{k}}-n^0_{2,\mathbf{k}}}{\tau_{\mathbf{k}}}-\Gamma^{out}_{2,\mathbf{k}} n_{2,\mathbf{k}}+\Gamma^{in}_{2,\mathbf{k}} (1-n_{2,\mathbf{k}})-i\sum_{\mathbf{q}}
\chi^{*}(\mathbf{q})Y(\mathbf{q,k-q})+c.c.\nonumber \\
\frac{d}{dt}Y(\mathbf{q,k})&=&
i(\omega_{c}(\mathbf{q})+\omega_{1}(\mathbf{k})-\omega_{2}(\mathbf{k+q})+i\Gamma_Y(\mathbf{q,k}))Y(\mathbf{q,k}) \\
&&-i\sum_{\mathbf{q'}}\chi(\mathbf{q})X(\mathbf{q+q',q',k})
+i\chi(\mathbf{q})n_{a,\mathbf{q}}
(n_{1,\mathbf{k}}-n_{2,\mathbf{k+q}})\nonumber\\
\frac{d}{dt}X(\mathbf{k'+q,k',k})&=& i(-\omega_{1}(\mathbf{k'})+\omega_{2}(\mathbf{k'+q})+\omega_{1}(\mathbf{k})-\omega_{2}(\mathbf{k+q}))X(\mathbf{k'+q,k',k})\nonumber\\
&&
- \Gamma_X(\mathbf{k'+q,k',k}) \left ( X(\mathbf{k'+q,k',k}) -2 n_{2,\mathbf{k+q}}(1-n_{1,\mathbf{k}})\delta_{\mathbf{k,k'}} \right )\nonumber \\
&&
+i\sum_{\mathbf{q'}}\chi(\mathbf{q''})
(Y^*(\mathbf{q',k})
\delta_{\mathbf{k'},\mathbf{k}}n_{2,\mathbf{k+q}}
+Y^*(\mathbf{q',q+k-q'})
\delta_{\mathbf{k'},\mathbf{k}}(1-n_{1,\mathbf{k}}))\nonumber\\
&& +2i\chi(\mathbf{q})Y^*(\mathbf{q,k'})
(n_{1,\mathbf{k}}-n_{2,\mathbf{k+q}}) -2i\chi^*(\mathbf{q})Y(\mathbf{q,k})
(n_{1,\mathbf{k'}}-n_{2,\mathbf{k'+q}})
\nonumber.\\ \nonumber
\end{eqnarray}
\begin{figure}[t!]
\begin{center}
\includegraphics[width=8cm]{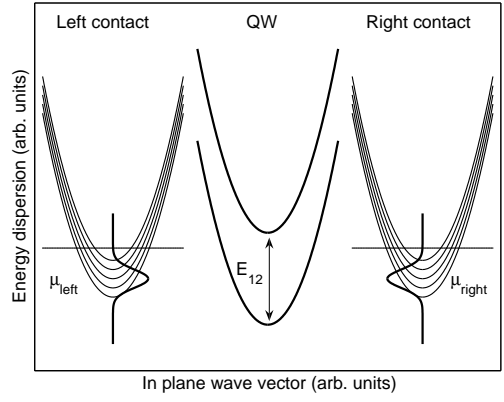}
\caption{ \label{nobias} Sketch of the energy dispersion of the
two quantum well subbands and of the minibands in the left and
right contacts in the zero-bias case. Here the system is in
thermal equilibrium and the Fermi level in the quantum well is the
same as in the two contacts. The doping level in the contacts
determines the equilibrium density in the quantum well. The
subband and minibands have an energy dispersion versus the
in-plane wavevector, which is a conserved quantity in the planar
structure. This electronic structure is embedded in a planar
microcavity, with a cavity photon mode quasi-resonant to the
intersubband transition.}
\end{center}
\end{figure}
\begin{figure}[t!]
\begin{center}
\includegraphics[width=8cm]{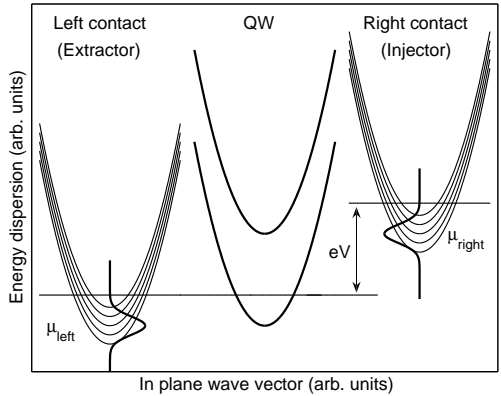}
\caption{\label{bias} Same as in Fig. \ref{nobias}, but with an
applied voltage bias. Here, the left contact acts as an electronic
extractor, while the right one is the injector. In the quantum
well, non-equilibrium steady-state populations can be established
in the two subbands. }
\end{center}
\end{figure}
\end{widetext}
\subsection{Injection and extraction tunneling rates}

The wave-vector dependent
injection and extraction rates in Eq. (\ref{full_system}) can be in principle of
different origin. Here we give the formal expression for elastic
tunneling processes conserving the in-plane momentum. Additional
processes (such as assisted tunneling) can be accounted for by
adding their contribution to the expressions for $\Gamma^{in}_{j,\mathbf{k}}$ and
$\Gamma^{out}_{j,\mathbf{k}}$ to be inserted in Eq. (\ref{full_system}).

As electronic contact reservoirs, we will consider semiconductor
doped superlattices, as it is generally the case in unipolar
quantum cascade devices.

The chemical
potential in each contact is labeled $\mu_{p}$ with $p={\rm left,right}$.
In each reservoir, we will consider miniband states with energy
$E^{res}_{p,\mathbf{k},k_z}$. In the elastic tunneling process,
electron energy and in-plane momentum are conserved. The tunneling
rate from the contact reservoir into the j-th subband is
\begin{eqnarray}
\Gamma_{p,j,\mathbf{k}}^{in}&=&\frac{2\pi}{\hbar}\sum_{k_z} \frac{\lvert
V_{p,j,\mathbf{k},k_z} \lvert ^2 \delta \left ( E^{res}_{p,\mathbf{k},k_z} - \hbar \omega_j(\mathbf{k}) \right )}{1+ e^{\beta (E^{res}_{p,\mathbf{k},k_z} - \mu_p)}},
\end{eqnarray}
where $V_{p,j,\mathbf{k},k_z}$ is the tunneling matrix element and $k_z$ is in general an index over the electronic states of the miniband with in-plane wave vector $\mathbf{k}$. It can be interpreted as the axial electronic wave vector in the case the two leads are just bulk contacts. $1/(1+ e^{\beta (E^{res}_{p,\mathbf{k},k_z} - \mu_p)})$ is the Fermi-Dirac occupation number of the electron states in the contact.
Analogously the tunneling rate from the j-th subband of the quantum well into the reservoir $p$ reads

\begin{eqnarray}
\Gamma_{p,j,\mathbf{k}}^{out}&=&\frac{2\pi}{\hbar}\sum_{k_z}\frac{ \lvert
V_{p,j,\mathbf{k},k_z} \lvert ^2 \delta \left ( E^{res}_{p,\mathbf{k},k_z} - \hbar \omega_j(\mathbf{k}) \right )}{1+ e^{-\beta (E^{res}_{p,\mathbf{k},k_z} - \mu_p)}} ,
\end{eqnarray}
where $1/(1+ e^{-\beta (E^{res}_{p,\mathbf{k},k_z} - \mu_p)}) = 1- 1/(1+ e^{\beta (E^{res}_{p,\mathbf{k},k_z} - \mu_p)})$ is the hole occupation number in the contact.
The value of $\Gamma_{p,j,\mathbf{k}}^{in,out}$ can be quantum engineered, depending on the specific structure. In particular, by changing the thickness of the potential barriers, it is possible to tailor considerably the tunneling matrix element. It is straightforward to see that a simple relationship occurs between $\Gamma_{p,j,\mathbf{k}}^{in}$  and $\Gamma_{p,j,\mathbf{k}}^{out}$, namely
\begin{equation}
\label{rapporto}
\frac{\Gamma_{p,j,\mathbf{k}}^{in}}{\Gamma_{p,j,\mathbf{k}}^{out}}=e^{\beta(\mu_p-\hbar \omega_j(\mathbf{k}))}.
\end{equation}
Note that here we have assumed that the bare energy dispersion of
the electrons in the two subbands is unaffected. This is valid in
the weak light-matter coupling regime or when the injector
miniband energy width is broad enough. For large values of the
vacuum Rabi frequency, the spectral function of the
electrons in the second subband is non-trivially modified as well
as the tunneling process using a narrow-band injector. This will
be proved and discussed in detail in a forthcoming paper\cite{Deliberato_exact}.

\section{Steady-state regime and observable quantities}
\label{steady} In this work, we will focus on the steady-state
solutions for the quantities $n_{a,\mathbf{q}}$, $n_{1,\mathbf{k}}$, $n_{2,\mathbf{k}}$,
$Y(\mathbf{q,k})$ and $X(\mathbf{q+q',q',k})$. Hence, we can set the time derivatives
equal to zero, transforming the differential system
(\ref{full_system}) into an algebraic one. In the steady-state
regime, the electronic current (number of electrons per unit time) through the structure is given by
the expression:
\begin{eqnarray}
I &=& \sum_{\mathbf{k}}  \Gamma^{out}_{1,\mathbf{k}} n_{1,\mathbf{k}} - \Gamma^{in}_{1,\mathbf{k}} (1-n_{1,\mathbf{k}}) \nonumber\\&=&  \sum_{\mathbf{k}} \Gamma^{in}_{2,\mathbf{k}} (1-n_{2,\mathbf{k}}) -\Gamma^{out}_{2,\mathbf{k}} n_{2,\mathbf{k}} .
\end{eqnarray}
The total rate of photons emitted out of the microcavity reads
\begin{equation}
P = 2 \gamma \sum_{\mathbf{q}}   n_{a,\mathbf{q}},
\end{equation}
where $1/(2 \gamma)$ is the escape time of a photon out of the
microcavity. The {\it quantum efficiency} $\eta$ is defined as the
ratio between the photonic current out of the cavity and
electronic current , i.e., $\eta = \frac{P}{I}$.

\section{Emission spectra}
\label{spectra} In the steady-state regime, the momentum-dependent
spontaneous photon emission spectra are given by the expression:
\begin{equation}
\mathcal{L}_{\mathbf{q}}(\omega) \propto \int_0^{\infty} dt  \Re
<a_{\mathbf{q}}^{\dagger}(0)a_{\mathbf{q}}(t)> e^{(i\omega -0^+)t} .
\end{equation}
In order to determine $<a_{\mathbf{q}}^{\dagger}(0)
a_{\mathbf{q}}(t)>$, we need to solve the following Heisenberg
equations of motion
\begin{widetext}
\begin{eqnarray}
\frac{d}{dt}<a_{\mathbf{q}}^{\dagger}(0)a_{\mathbf{q}}>&=&
-i\omega_{c}(\mathbf{q})<a_{\mathbf{q}}^{\dagger}(0)a_{\mathbf{q}}>+i
\chi^{*}(\mathbf{q})\sum_{\mathbf{k},\sigma}<a_{\mathbf{q}}^{\dagger}(0)c_{1,\sigma,\mathbf{k}}^{\dagger}c_{2,\sigma,\mathbf{k+q}}>\\
\frac{d}{dt} <a_{\mathbf{q}}^{\dagger}(0)c_{1,\sigma,\mathbf{k}}^{\dagger}c_{2,\sigma,\mathbf{k+q}}>&=&-i\omega_{12}<a_{\mathbf{q}}^{\dagger}(0)c_{1,\sigma,\mathbf{k}}^{\dagger}c_{2,\sigma,\mathbf{k+q}}>\nonumber \\&&-i\sum_{\mathbf{q'}}
\chi(\mathbf{q'})<a_{\mathbf{q}}^{\dagger}(0)a_{\mathbf{q'}}c_{2,\sigma,\mathbf{k+q'}}^{\dagger}c_{2,\sigma,\mathbf{k+q}}>
+i\sum_{\mathbf{q'}}
\chi(\mathbf{q'})<a_{\mathbf{q}}^{\dagger}(0)a_{\mathbf{q'}}c_{1,\sigma,\mathbf{k}}^{\dagger}c_{1,\sigma,\mathbf{k+q-q'}}>\nonumber.
\end{eqnarray}

Note that here we have omitted the coupling of the electronic
injector and extractor reservoirs to the quantity
$<a_{\mathbf{q}}^{\dagger}(0)c_{1,\sigma,\mathbf{k}}^{\dagger}c_{2,\sigma,\mathbf{k+q}}>$.
This coupling would involve correlations between the quantum well
electronic field and the contact electronic fields. Since in this
paper we are dealing with incoherent electron transport, we will
neglect such correlations with the contact reservoirs, which are
also extremely tricky to tackle.

Truncating the hierarchy at the level of two excitations (details
in Appendix \ref{algebraic}) and taking the unilateral Fourier transform
($\int_{0}^{\infty} dt e^{i\omega t}$) we obtain:

\begin{eqnarray}
S_{\mathbf{q}}(t=0)&=&n_{a,\mathbf{q}}=
i(\omega-\omega_{c}(\mathbf{q})+i\Gamma_S(\mathbf{q}))\tilde{S}_{\mathbf{q}}(\omega)+2i
\chi^{*}(\mathbf{q})\tilde{Z}_{\mathbf{q}}(\omega)\\
Z_{\mathbf{q}}(t=0)&=&\sum_{\mathbf{k}}Y(\mathbf{q},\mathbf{k})=i(\omega-\omega_{12}+i\Gamma_Z(\mathbf{q}))\tilde{Z}_{\mathbf{q}}(\omega)
+i\chi(\mathbf{q})\tilde{S}_{\mathbf{q}}(\omega)D.\nonumber
\end{eqnarray}
where $S_{\mathbf{q}}(t)=<a_{\mathbf{q}}^{\dagger}(0)a_{\mathbf{q}}(t)>$,
$Z_{\mathbf{q}}(t)=\sum_{\mathbf{k}}<a_{\mathbf{q}}^{\dagger}(0)c_{1,\sigma,\mathbf{k}}^{\dagger}c_{2,\sigma,\mathbf{k+q}}>$ and $D$ represents half the difference between the total number of electrons in the fundamental subband and the
number in the second one, namely:
\begin{equation}
D = \sum_{\mathbf{k}} D_{\mathbf{k}} = \sum_{\mathbf{k}} n_{1,\mathbf{k}}-n_{2,\mathbf{k}}.
\end{equation}
Note that the total density of electrons is  $2 \sum_{\mathbf{k}} n_{1,\mathbf{k}} + n_{2,\mathbf{k}}$, where the $2$ factor accounts for the two-fold spin degeneracy of the electron states in the conduction subbands. $\Gamma_{S}$ and $\Gamma_{Z}$ are phenomenological damping rates for $S_{\mathbf{q}}$ and $Z_{\mathbf{q}}$ respectively.
The analytical solutions are
\begin{eqnarray}
\label{spectrum_formula} \tilde{S}_{\mathbf{q}}(\omega)&=&\frac{
 in_{a,\mathbf{q}}\left ( \gamma
(\frac{\omega_{c}(\mathbf{q})-\omega_{12}}{\Gamma_Y}+i)-(\omega-\omega_{12}+i\Gamma_Z)
\right
)}{(\omega-\omega_{12}+i\Gamma_Z)(\omega-\omega_{c}(\mathbf{q})+i\Gamma_S)-2\chi(\mathbf{q})^2{D}},
\label{EL}
\end{eqnarray}
\begin{eqnarray*}
\tilde{Z}_{\mathbf{q}}(\omega)&=&
-\frac{\chi(\mathbf{q})S_{\mathbf{q}}(\omega)D+i\frac{\gamma
n_{a,\mathbf{q}}}{2\chi(\mathbf{q})}
(\frac{\omega_{c}(\mathbf{q})-\omega_{12}}{\Gamma_Y}-i)
}{\omega-\omega_{12}+i\Gamma_Z}.
\end{eqnarray*}
\end{widetext}

The electroluminescence spectrum is simply
\begin{equation}
\mathcal{L}_{\mathbf{q}}(\omega) \propto  \Re
\tilde{S}_{\mathbf{q}}(\omega) .
\end{equation}
From the analytical result for $\tilde{S}_{\mathbf{q}}(\omega)$,
we see immediately that emission spectrum is resonant at the two
polariton frequencies $\omega_{\pm}(\mathbf{q})$ satisfying the
equation
\begin{equation}
(\omega-\omega_{12}+i\Gamma_Z)(\omega-\omega_{c}(\mathbf{q})+i\Gamma_S)-2\chi(\mathbf{q})^2D = 0.
\end{equation}
The quantity $\Omega_R = \chi(\mathbf{q})\sqrt{2D}$ is just
the vacuum Rabi frequency of the present system. At resonance
(i.e., $\omega_{c}(\mathbf{q}) = \omega_{12}$),  the
necessary condition for the appearance of a strong coupling
polaritonic splitting is  $D > D_0 = \frac{(\Gamma_S-\Gamma_Z)^2}{8\chi(\mathbf{q})^2} $, meaning that
the total density of electrons in the fundamental subband must be
larger enough than the total density in the second. For a vacuum
Rabi frequency much larger than $\Gamma_Z$ and $\Gamma_S$, the
minimum polariton splitting is given by twice the vacuum Rabi
frequency.

Note that here the electroluminescence spectral shape does not
depend explicitly on the spectral properties of the injector and
extractor reservoirs. The spectrum in Eq. (\ref{EL}) has the same
shape as the absorption (in presence of the same carrier
densities). The dependence on the transport is only implicit,
being given by the steady-state carrier and photon populations. In
contrast, in the exact solution of the the simplified model of
Ref. \onlinecite{Ciuti_PRA}, it is shown that the
electroluminescence spectra is the absorption spectrum times the
spectral distribution of excitations in the electronic reservoir,
which then acts as an electronic filter\cite{Luca-unpublished}. A
fermionic approach based on an exact diagonalization method
\cite{Deliberato_exact} indeed shows that the spectral properties
of the electronic contact modifies significantly the spectral
shape of the electroluminescence in the case of narrow band
injectors. Hence, the spectrum predicted by Eq. (\ref{EL}) is
valid only for broad band injectors. This is not really surprising
because, in order to calculate the tunneling rates, we have
considered bare electronic states in the quantum well and have
only considered incoherent population injection and extraction processes.

\section{Numerical application}
\begin{figure}[t!]
\begin{center}
\includegraphics[width=8cm]{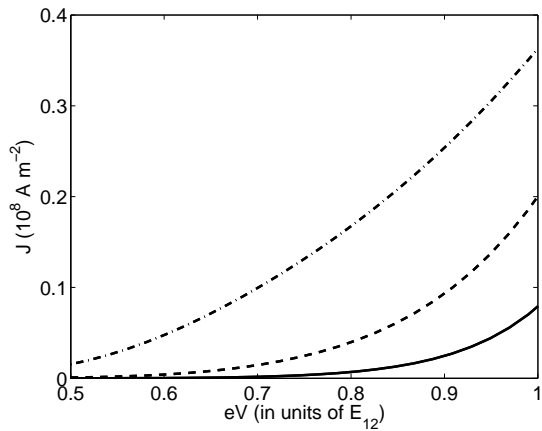}
\caption{\label{IV-curve} Current density
versus applied voltage for different values of the intersubband
transition energy: $E_{12}=50$meV, (dashed-dotted
line), $100$ meV (dashed line) and $150$meV (solid line). Other parameters
can be found in the text.}

\end{center}
\end{figure}
\begin{figure}[t!]
\begin{center}
\includegraphics[width=8cm]{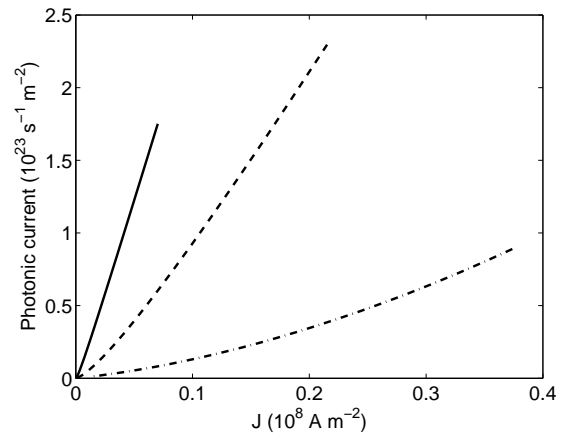}
\caption{\label{photons} Photonic current density versus
electronic current for different values of the intersubband
transition energy: $E_{12}=50 $meV, (dashed-dotted
line), $100 $meV (dashed line) and $150$meV (solid line). Same
parameters and range of applied voltages as in Fig.
\ref{IV-curve}.}
\end{center}
\end{figure}

\begin{figure}[t!]
\begin{center}
\includegraphics[width=8cm]{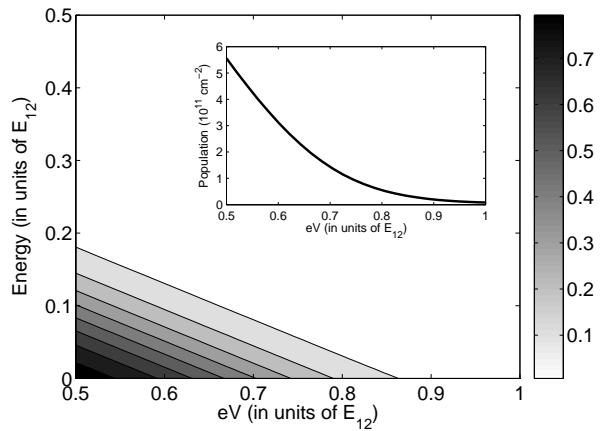}
\caption{\label{n1}Electron occupation number in the fundamental
conduction subband as a function of kinetic energy and applied
voltage. Inset: the integrated density of electrons in the
fundamental subband versus voltage. $E_{12}=150 $meV
and other parameters as in Fig. \ref{IV-curve}.  For $eV=E_{12}$, the density of electrons in the first subband is $8.3\times 10^9 {\rm cm}^{-2}$. }
\end{center}
\end{figure}

\begin{figure}[b!]
\begin{center}
\includegraphics[width=8cm]{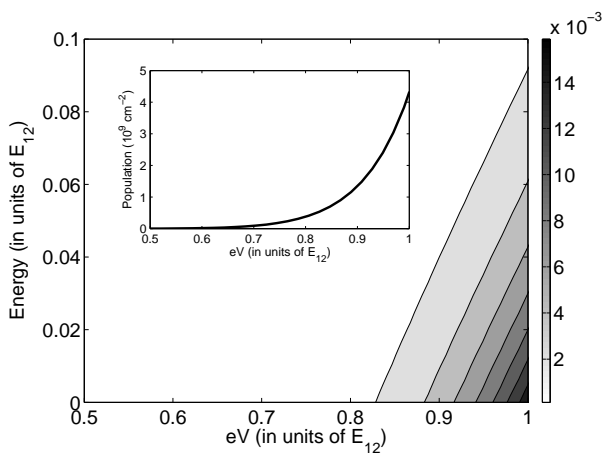}
\caption{\label{n2} Same as in Fig. \ref{n1}, but for the second
subband. Inset: the integrated density of electrons in the second
subband versus voltage. For $eV=E_{12}$, the density of electrons in the second subband is $4.3 \times 10^9 {\rm cm}^{-2}$. }
\end{center}
\end{figure}

\begin{figure}[b!]
\begin{center}
\includegraphics[width=8cm]{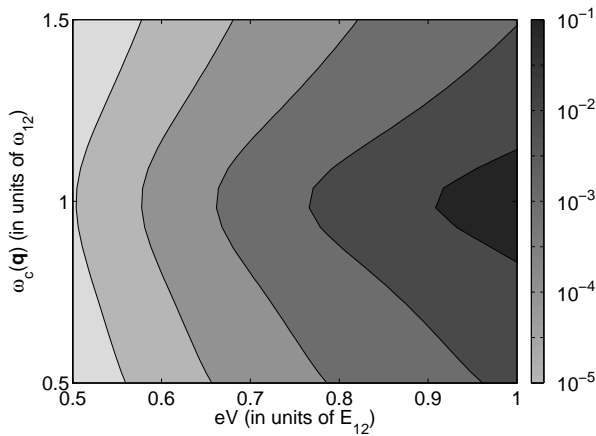}
\caption{\label{photon_k}Contour plot of the photon occupation
(log scale) versus the applied voltage and the energy of the bare cavity photon mode. }
\end{center}
\end{figure}
\label{numerical}

Here, we apply our theory using realistic parameters for a
microcavity-embedded quantum cascade electroluminescent source. In
order to simplify the algebra, we have systematically neglected
the photon wavevector whenever added to an electronic wavevector.
Given the huge difference in the typical wavectors of photons and
electrons, this simplification is safe. Applying this
approximation, we can obtain a closed set of algebraic equation
where the variables are the populations in the two subbands and in
the cavity photonic branch, as shown in Appendix \ref{algebraic}.
This system has been solved numerically using a standard Newton
method. We achieve numerical convergence in a relatively fast
computation time except in the limit of vanishing bias, when the
injector and extractor are strongly 'misaligned' with the two
subbands. Physically in this case the steady-state situation is
reached in times very long compared to the dynamics of the quantum
well system, the photon population is extremely small and
correspondingly the numerical method fails to converge. Anyway
this is not a real limitation, because we are interested in the
behavior of the system in presence of a finite voltage bias,
producing a significant current flow and photonic output.

In Figs. \ref{nobias} and \ref{bias}, we show a
sketch of the energy profile of the injector and extractor with
respect to the quantum well subbands respectively without and with
an applied bias. Specifically, in the numerical calculations we
have used the following electronic out-tunneling rates:
\begin{eqnarray}
\label{injector}
\Gamma_{{\rm left},j,\mathbf{k}}^{out}&=& \frac{\Gamma e^{-\frac{(E_{0,{\rm left}}-qV/2)^2}{2\sigma^2}} }{1+ e^{\beta (-\hbar \omega_{j,\mathbf{k}} + \mu_{{\rm left}}-qV/2)}}, \\
\Gamma_{{\rm right},j,\mathbf{k}}^{out}&=& \frac{\Gamma
e^{-\frac{(E_{0,{\rm right}}+qV/2)^2}{2\sigma^2}} }{1+ e^{\beta
(-\hbar \omega_{j,\mathbf{k}} + \mu_{{\rm
right}}+qV/2)}},\nonumber
\end{eqnarray}
where $\sigma=0.1E_{12}$, $1/\Gamma=0.4$ps, $E_{0,{\rm
left}}$ and $E_{0,{\rm right}}$ are the energy offsets of the left
and right minibands. The in-tunneling rates are determined by
applying the relation in Eq. (\ref{rapporto}). In all the
simulations, we have taken $E_{0,{\rm left}}=E_{0,{\rm
right}}=0.5\hbar\omega_{12}$ and $\mu_{{\rm left}}=\mu_{{\rm
right}}=\frac{1}{3}\hbar\omega_{12}$.

Note that these are
just phenomenological injection rates. For the amplitude $\Gamma$, we
have considered values which are consistent with what realistically
obtainable in semiconductor intersubband devices. Importantly, in real
structures $\Gamma$ can be considerably quantum engineered by changing
the barrier thickness and/or the miniband structure of the injection
superlattices. This is why we have not considered a very specific
injector configuration and taken the simplified expression in Eq. (\ref{injector})
with realistic parameters.

When a voltage bias is applied, the two reservoirs are shifted
symmetrically, as shown in Fig.\ref{bias}. In all the simulations,
except when otherwise stated we used the realistic damping
parameters  $\Gamma_X=\Gamma_Y= \Gamma_S=\Gamma_Z=0.1\omega_{12}$,
$\gamma=0.05\omega_{12}$, while the temperature is $T=77$K. In the
simulations we have also considered $\tau_{\mathbf{k}}$ to be
independent from $\mathbf{k}$ and such that
$\frac{1}{\tau}=0.005\omega_{12}$, except when otherwise
stated. Note that here we have considered only an active quantum
well. For quantum cascade structures with several active quantum
wells repeated in a periodic way, the dynamics is similar and the
present treatment can be generalized without major difficulties.
In the simulations, the intersubband transition energy
$E_{12}=\hbar\omega_{12}$ is, except where otherwise stated, equal
to $150$meV and the coupling constant $\chi(\mathbf{q})$ is such
that the vacuum Rabi frequency is $0.1\omega_{12}$ for an electron
density of $5\times 10^{11}$cm$^{-2}$ (all in the fundamental
subband). When $E_{12}$ is changed, the coupling constant is
adjusted in order to keep the ratio between the vacuum Rabi
frequency and transition frequency constant. The effective mass
$m^*$ has been taken to be one tenth of the bare electronic mass.
In the numerical calculations, the cavity spacer dielectric
constant is $\epsilon_r = 10$. For each simulation, the resonance
in-plane wavector $q_{res}$, given by the condition
$\omega_{c}(q_{res}) = \omega_{12}$, corresponds to an internal
cavity photon propagation angle $\theta_{res}$ equal to 70
degrees, where $\tan \theta_{res}= q_{res}/q_z$.

In Fig. \ref{IV-curve}, we show the current density versus applied
voltage (between the injector and extractor) for different values
of $E_{12}$. The current-voltage profile is characteristic of an
unipolar quantum cascade light emitting diode. The current grows
superlinearly in the voltage region where the injector Fermi level
approaches the second subband. The current is bigger for smaller
$E_{12}$ because, keeping the injection rate $\Gamma$ constant
(but all the internal rates of the system proportional to
$E_{12}$), the injection and extraction processes become the
dominant processes. Note that an increase of the nonradiative
relaxation rate $1/\tau$ produces a nearly proportional increase
of the electronic current (not shown). The rates of emitted
photons per unit area (integrated all over the in-plane
wavevectors) are shown in Fig. \ref{photons} as a function of the
flowing current, showing an approximately linear behavior.

Fig. \ref{n1} and \ref{n2} show contour plots of the electron
occupation numbers of the first and second subband respectively as
a function of the applied voltage and of the kinetic energy. The
insets in Fig. \ref{n1} and \ref{n2} show respectively the
integrated density of electrons in the first and second subband.
It is apparent that with increasing voltage the population in the
first subband decreases, while the population in the second
subband increases.

When the injector Fermi level becomes
aligned with the second subband, as expected, the carrier occupation
numbers in the two subbands are considerably out of equilibrium. The
decrease of the first subband carrier occupation numbers is beneficial
for the radiative efficiency of the spontaneous emission, because the
influence of Pauli blocking is reduced. Moreover, in the considered
conditions, the density of electrons in the first subband is still
considerably larger than in the second subband, thus producing a large
vacuum Rabi coupling and efficient emission rate.

\begin{figure}[t]
\begin{center}
\includegraphics[width=9cm]{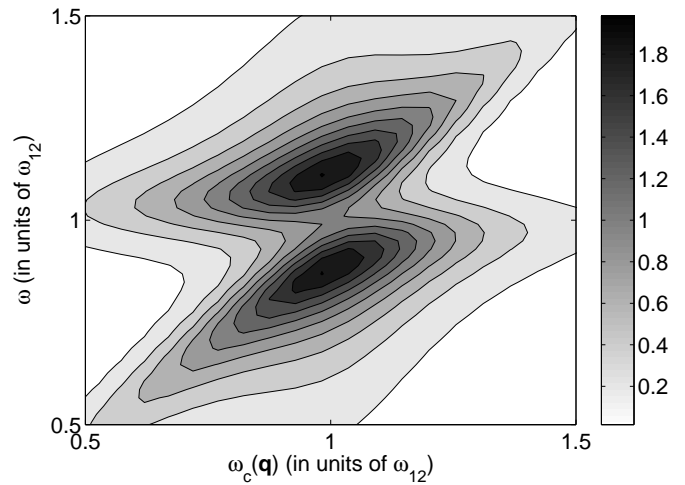}
\caption{\label{spectrum} Contour plot of the electroluminescence
(arb. units) as a function of the bare cavity photon energy
$\omega_c(q)$ and of the emission frequency $\omega$ for an
applied voltage $eV=0.5E_{12}$. The anticrossing of the two
intersubband polariton branches is apparent in electroluminescence
spectra.}
\end{center}
\end{figure}

\begin{figure}[t]
\begin{center}
\includegraphics[width=9cm]{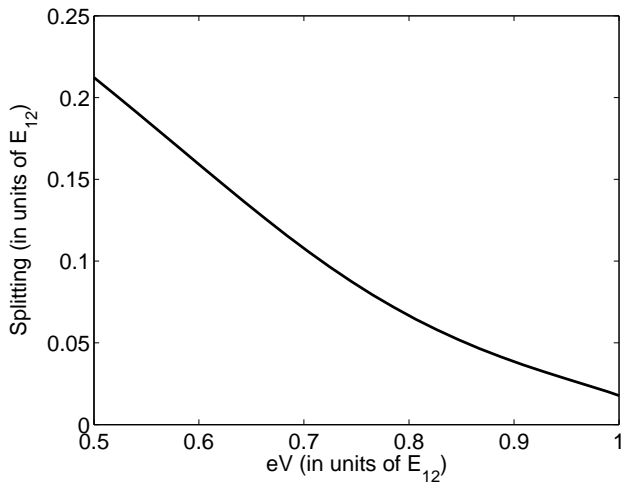}
\caption{\label{splitting} Minimum polariton splitting as a
function of the applied voltage. $E_{12}=150 $meV and other
parameters can be found in the text.}
\end{center}
\end{figure}

 Fig. \ref{photon_k} contains a contour plot of
the cavity photon occupation number versus the bare photon energy,
showing that the maximum of emission is obtained when the bare photon energy is resonant with the
intersubband transition, as expected and as observed
experimentally \cite{Luca-unpublished,Luca_thesis}. With the
considered parameters, the density of electrons in the first
subband is high enough to be in the strong
coupling regime, as depicted in Fig. \ref{spectrum}, where the
anticrossing of two polariton branches is clearly present. The
minimum polariton splitting, given by the expression $2 \chi(q)
\sqrt{2D}$ is reported in Fig. \ref{splitting} as a function of
the applied bias. With increasing voltage, the population
difference $D = \sum_{\mathbf{k}} D_{\mathbf{k}} =
\sum_{\mathbf{k}} n_{1,\mathbf{k}}-n_{2,\mathbf{k}}$ diminishes.
This results in a decrease of the vacuum Rabi frequency and
consequently of the polariton splitting. This high-excitation
feature has been already observed in
experiments\cite{Luca-unpublished,Luca_thesis} and can not be
described within a bosonic approach\cite{Ciuti_PRA}, which can be
applied only to the low excitation density case\cite{pereira}.

It is interesting to analyze the quantum efficiency $\eta$,
defined as the ratio between the photonic emission rate and the
electronic current, namely
\begin{equation}
\eta =  \frac{2 \gamma
\sum_{\mathbf{q}}n_{a,\mathbf{q}}}{{{\sum_{\mathbf{k}}}}
\Gamma^{out}_{1,\mathbf{k}} n_{1,\mathbf{k}} -
\Gamma^{in}_{1,\mathbf{k}} (1-n_{1,\mathbf{k}})} .
\end{equation}

\begin{figure}[t]
\begin{center}
\includegraphics[width=9.3cm]{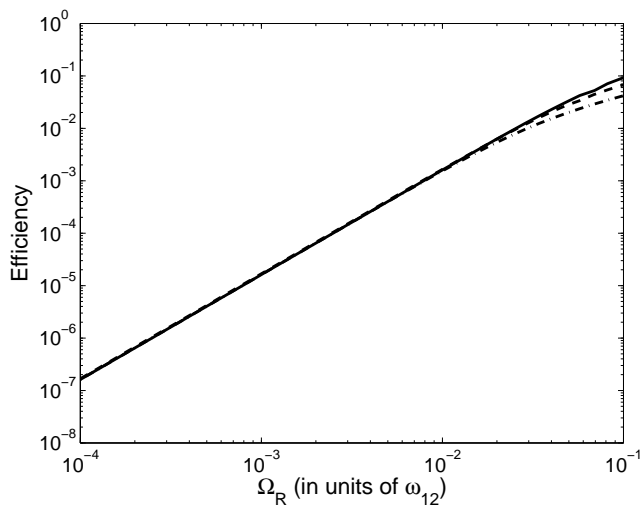}
\caption{\label{efficiencyGamma} Quantum efficiency versus the
corresponding vacuum Rabi frequency at the voltage $eV = E_{12}$.
The three lines are obtained with different values of the
coherence damping coefficients: $\Gamma_X=\Gamma_Y=0.1\omega_{12}$
(solid line), $0.05\omega_{12}$ (dashed line) and
$0.025\omega_{12}$ (dashed-dotted line). $E_{12}=150 $meV.}
\end{center}
\end{figure}

In Fig. \ref{efficiencyGamma}, we plot the quantum efficiency
$\eta$ at $eV=E_{12}$ versus the vacuum Rabi frequency
$\Omega_{R}$ at the same voltage (log-log scale). In the simulations, 
the vacuum Rabi frequency has been varied by changing the coupling constant
$\chi({\mathbf{q}})$. In a realistic quantum engineered device, $\chi({\mathbf{q}})$ can be 
tailored in different ways. For example, by growing the active quantum wells in a spatial region
where the cavity mode field is very small, it is possible to quench dramatically the value of
$\chi({\mathbf{q}})$. Moreover, by using different shape of quantum wells, it is also possible
to tailor the transition dipole $d_{12}$. Fig. \ref{efficiencyGamma} shows that in the weak coupling
regime (small values of $\Omega_R$) the efficiency grows like
$\Omega_R^2$. In the strong coupling regime, the efficiency
becomes impressive and then tends to saturate.
It is apparent that the radiative efficiency smoothly increases passing from the weak to the strong coupling regime. This crossover occurs because the radiative efficiency depends on the spectrally integrated emission and it is therefore insensitive to the sudden appearence of the polariton doublet in the strong coupling emission spectra.

\begin{figure}[t]
\begin{center}
\includegraphics[width=9.3cm]{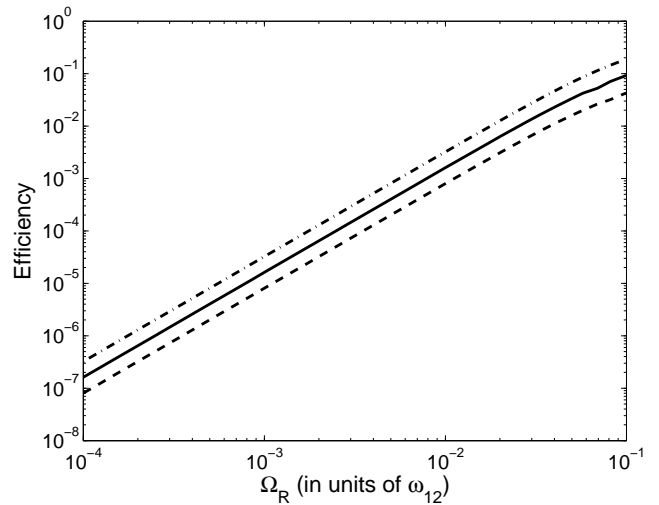}
\caption{\label{efficiencytau} Quantum efficiency versus the
corresponding vacuum Rabi frequency at the voltage $eV = E_{12}$.
The three lines are obtained with different values of the
non-radiative relaxation rate: $\frac{1}{\tau}=0.01\omega_{12}$
(dashed line), $0.005\omega_{12}$ (solid line) and
$0.0025\omega_{12}$ (dashed-dotted line). $E_{12}=150 $meV. }
\end{center}
\end{figure}

 This results are
in qualitative agreement with the analytical solutions of the
simplified model in Ref. \onlinecite{Ciuti_PRA}, where only the
bright intersubband states are considered and where the electronic
reservoir is modeled with a bath of harmonic oscillators. As shown
in Fig. \ref{efficiencytau}, the nonradiative population
relaxation rate $1/\tau$ has the most significant effect. In the considered regime of parameters, the
efficiency is proportional to $\tau$.

It is interesting to compare our results for this microcavity system
with the standard free space case. In the free-space case, the photon
current, obtained by applying the Fermi golden rule, is given by the
formula $P=\frac{2 d_{12}^2\omega_{12}^3\sqrt{\epsilon_r} }{3\pi c^3 \hbar \epsilon_0} \sum_{\mathbf{k}} n_{2,\mathbf{k}}(1-n_{1,\mathbf{k}})$. 
As it is well known, the free-space radiative efficiency
dramatically decreases with the intersubband emission wavelength due to
the $\omega_{12}^3 d_{12}^2$ dependence of the spontaneous emission rate
($d_{12}^2 \propto 1/\omega_{12}$, so the spontaneous emission rate
scales effectively as $\omega_{12}^2$). In the mid-infrared, by using the
same parameters, for a transition of $150$ meV, the quantum efficiency is
of the order of $10^{-4}-10^{-5}$. Hence, it is clear from our results that a
strong coupling light-emitting diode based on a planar microcavity
system can provide a dramatic enhancement with respect to the free space
case (even three orders of magnitude for the larger vacuum Rabi
frequency case).

\section{Conclusions and perspectives}
\label{conclusions} In conclusion, we have presented a quantum
theoretical study of the quantum well intersubband
electroluminescence from a semiconductor microcavity in the
incoherent transport regime, i.e. when the coupling to the
electronic contacts concerns only the electron populations. The
problem has been tackled starting from the fermionic electron
Hamiltonian for the two-subbands and by using a cluster
factorization method to truncate the infinite hierarchy of
dynamical equations for the relevant expectation values of
operator products. At the present level of approximation, we have
been able to describe the incoherent electron transport through
the dynamics of the electronic subband populations and the
electroluminescence through the dynamics of the cavity photon
population, which is coupled to the correlations between the
electromagnetic field and the intersubband polarization. We have
discussed the limits of applicability of the present approach,
which neglects the impact of the vacuum Rabi coupling on the quantum well
carrier spectral function and any correlation between the quantum
well and the contact reservoirs. The analogies and differences
with the exact predictions of the simplified model in Ref.
\onlinecite{Ciuti_PRA} have been critically and extensively discussed.
We have shown the appearance of cavity polariton resonances in the
emission spectra, when a large density of electrons occupies the
fundamental subband. We have described how the vacuum Rabi
splitting decreases with increasing voltage and described the
photonic output in the different transport conditions. Our results
show that even in presence of non-radiative relaxation and Pauli
blocking, the quantum efficiency of the microcavity intersubband
electroluminescence can be considerably enhanced by increasing the
vacuum Rabi frequency. A more refined
treatment\cite{Deliberato_exact} based on a fermionic exact
diagonalization method shows that under certain conditions the
strong vacuum Rabi coupling regime affect considerably not only
the dynamics of the intersubband polarization (hence the
absorption spectrum), but also the quantum well electron spectral
properties and consequently the tunneling transport using
narrow-band injectors. As future perspective, this could be
exploited to further improve the quantum efficiency of microcavity
intersubband emitters and for the eventual realization of intersubband polariton lasers.

\begin{acknowledgements}
We are pleased to thank I. Carusotto, R. Colombelli, L. Sapienza,
C. Sirtori, A. Vasanelli for  discussions.

\end{acknowledgements}
\appendix
\section{Factorizations}
\label{app}
\begin{widetext}

As stated in the main body of the paper we used a cluster
expansion and truncation scheme to obtain a closed set of
equations. Here we briefly review the principles of this method
following \cite{F1,F2,KK1} and apply it to the actual case.

If we consider each bosonic operator or each pair of fermionic
operators as an excitation operator and we write the expectation
value of an $N$ excitation operator as $<N>$, then the Heisenberg
equation of motion takes the form:

\begin{equation*}
i\frac{\partial}{\partial t}<N>=T\lbrack <N>\rbrack+V\lbrack <N+1>\rbrack
\end{equation*}

where the N-excitation expectation value is coupled to higher
order quantities via the functional V. An N-excitation truncation
scheme is obtained if we factorize all the expectation values of
more than $N$ excitation in all the possible ways and considering
the sign exchange for the fermionic operators in order to obtain a
factorized quantity that respects the commutation and
anticommutation properties of the original quantity.

We are interested in incoherent emission only, so the only
nonzero one excitation operators we consider are
$<c_{1,\sigma,\mathbf{k}}^{\dagger}c_{1,\sigma,\mathbf{k}}>$ and
$<c_{2,\sigma,\mathbf{k}}^{\dagger}c_{2,\sigma,\mathbf{k}}>$. We
factorized the $3$ excitations operators in the following way:

\begin{eqnarray*}
<a_{\mathbf{q}}c_{1,\sigma,\mathbf{k}}c_{2,\sigma,\mathbf{k'}}^{\dagger}c_{2,\sigma',\mathbf{k''}}^{\dagger}c_{2,\sigma',\mathbf{k'''}}>&=&-<a_{\mathbf{q}}c_{2,\sigma,\mathbf{k'}}^{\dagger}c_{1,\sigma,\mathbf{k}}><c_{2,\sigma',\mathbf{k''}}^{\dagger}c_{2,\sigma',\mathbf{k'''}}>+<a_{\mathbf{q}}c_{2,\sigma',\mathbf{k''}}^{\dagger}c_{1,\sigma,\mathbf{k}}><c_{2,\sigma,\mathbf{k'}}^{\dagger}c_{2,\sigma',\mathbf{k'''}}>\\
&=&
-<a_{\mathbf{q}}c_{2,\sigma,\mathbf{k'}}^{\dagger}c_{1,\sigma,\mathbf{k}}>
\delta_{\mathbf{k''},\mathbf{k'''}}n_{2,\mathbf{k''}}
+<a_{\mathbf{q}}c_{2,\sigma,\mathbf{k''}}^{\dagger}c_{1,\sigma,\mathbf{k}}>
\delta_{\mathbf{k'},\mathbf{k'''}}\delta_{\mathbf{\sigma},\mathbf{\sigma'}}n_{2,\mathbf{k'}}\\
<a_{\mathbf{q}}c_{2,\sigma,\mathbf{k}}^{\dagger}c_{1,\sigma,\mathbf{k'}}c_{1,\sigma',\mathbf{k''}}c_{1,\sigma',\mathbf{k'''}}^{\dagger}>
&=&
-<a_{\mathbf{q}}c_{2,\sigma,\mathbf{k}}^{\dagger}c_{1,\sigma',\mathbf{k''}}><c_{1,\sigma,\mathbf{k'}}c_{1,\sigma',\mathbf{k'''}}^{\dagger}>
+<a_{\mathbf{q}}c_{2,\sigma,\mathbf{k}}^{\dagger}c_{1,\sigma,\mathbf{k'}}><c_{1,\sigma',\mathbf{k''}}c_{1,\sigma',\mathbf{k'''}}^{\dagger}>\\
&=&
-<a_{\mathbf{q}}c_{2,\sigma,\mathbf{k}}^{\dagger}c_{1,\sigma,\mathbf{k''}}>
\delta_{\mathbf{k'},\mathbf{k'''}}\delta_{\mathbf{\sigma},\mathbf{\sigma'}}(1-n_{1,\mathbf{k'}})
+<a_{\mathbf{q}}c_{2,\sigma,\mathbf{k}}^{\dagger}c_{1,\sigma,\mathbf{k'}}>
\delta_{\mathbf{k''},\mathbf{k'''}}(1-n_{1,\mathbf{k''}})\\
\end{eqnarray*}

For the two-time quantities in the calculation of luminescence
spectrum, we proceed analogously and obtain:

\begin{eqnarray*}
<a_{\mathbf{q}}^{\dagger}(0)a_{\mathbf{q'}}c_{2,\sigma,\mathbf{k+q}}^{\dagger}c_{2,\sigma,\mathbf{k+q'}}>&=&<a_{\mathbf{q}}^{\dagger}(0)a_{\mathbf{q'}}><c_{2,\sigma,\mathbf{k+q}}^{\dagger}c_{2,\sigma,\mathbf{k+q'}}>\delta_{\mathbf{q},\mathbf{q'}}, \\<a_{\mathbf{q}}^{\dagger}(0)a_{\mathbf{q'}}c_{1,\sigma,\mathbf{k}}^{\dagger}c_{1,\sigma,\mathbf{k+q-q'}}>&=&<a_{\mathbf{q}}^{\dagger}(0)a_{\mathbf{q'}}><c_{1,\sigma,\mathbf{k}}^{\dagger}c_{1,\sigma,\mathbf{k+q-q'}}>\delta_{\mathbf{q},\mathbf{q'}}.
\end{eqnarray*}

\section{Algebraic equations for the steady-state regime}
\label{algebraic} In the steady-state regime, neglecting
the photonic wavevector into sums over electronic wavevectors, the
system of equations (\ref{full_system}) reduces to the following
system of algebraic equations.

\begin{eqnarray*}
0&=&\left (B_{\mathbf{q}}(\gamma +\Gamma_X)+(\frac{\delta_{\mathbf{q}}^2}{\Gamma_Y}+\frac{G_{\mathbf{q}}\Gamma_X}{2D\chi(\mathbf{q})^2})
\gamma\right )n_{a,\mathbf{q}}+\frac{B_{\mathbf{q}}}{D}\sum_{\mathbf{k}}(1-D_{\mathbf{k}})
(\frac{n_{1,\mathbf{k}}-n^0_{1,\mathbf{k}}}{\tau_{\mathbf{k}}} +\Gamma^{out}_{1,\mathbf{k}} n_{1,\mathbf{k}}-\Gamma^{in}_{1,\mathbf{k}} (1-n_{1,\mathbf{k}}))
-\frac{2B_{\mathbf{q}}F\Gamma_X}{D},
\end{eqnarray*}
\begin{eqnarray*}
0&=&(\sum_{\mathbf{q}}\frac{B_{\mathbf{q}}\chi(\mathbf{q})^2}{G_{\mathbf{q}}\Gamma_X}(1-D_{\mathbf{k}})+\frac{1}{2})
(\frac{n_{1,\mathbf{k}}-n^0_{1,\mathbf{k}}}{\tau_{\mathbf{k}}} +\Gamma^{out}_{1,\mathbf{k}} n_{1,\mathbf{k}}-\Gamma^{in}_{1,\mathbf{k}} (1-n_{1,\mathbf{k}}))
\\&&
+\frac{D_{\mathbf{k}}}{\Gamma_X\Gamma_Y}\sum_{\mathbf{q}}\frac{\chi(\mathbf{q})^2n_{a,\mathbf{q}}}{G_{\mathbf{q}}}(\Gamma_Y B_{\mathbf{q}}(\gamma +\Gamma_X)+\delta_{\mathbf{q}}^2\gamma)-2F_{\mathbf{k}}\sum_{\mathbf{q}}\frac{B_{\mathbf{q}}\chi(\mathbf{q})^2}{G_{\mathbf{q}}},
\end{eqnarray*}
\begin{eqnarray*}
0&=&-\frac{n_{2,\mathbf{k}}-n^0_{2,\mathbf{k}}}{\tau_{\mathbf{k}}}-\Gamma^{out}_{2,\mathbf{k}} n_{2,\mathbf{k}}+\Gamma^{in}_{2,\mathbf{k}} (1-n_{2,\mathbf{k}})-\frac{n_{1,\mathbf{k}}-n^0_{1,\mathbf{k}}}{\tau_{\mathbf{k}}} -\Gamma^{out}_{1,\mathbf{k}} n_{1,\mathbf{k}}+\Gamma^{in}_{1,\mathbf{k}} (1-n_{1,\mathbf{k}}),
\end{eqnarray*}
where
\begin{eqnarray*}
D_{\mathbf{k}}&=&n_{1,\mathbf{k}}-n_{2,\mathbf{k}}\\
F_{\mathbf{k}}&=&n_{2,\mathbf{k}}(1-n_{1,\mathbf{k}})\\
D&=&\sum_{\mathbf{k}} D_{\mathbf{k}}\\
F&=&\sum_{\mathbf{k}} F_{\mathbf{k}}\\
n^0_{1,\mathbf{k}}&=&\frac{1}{\exp{\beta(\omega_1(\mathbf{k})-\epsilon_F)}+1}\\
n^0_{2,\mathbf{k}}&=&\frac{1}{\exp{\beta(\omega_2(\mathbf{k})-\epsilon_F)}+1}\\
B_{\mathbf{q}}&=&\Gamma_Y+\frac{2\chi(\mathbf{q})^2}{\Gamma_X}D\\
\delta_{\mathbf{q}}&=&\omega_{c}(\mathbf{q})-\omega_{12}\\
G_{\mathbf{q}}&=&(\omega_{c}(\mathbf{q})-\omega_{12})^2+(\Gamma_Y+\frac{2\chi(\mathbf{q})^2D}{\Gamma_X})^2.
\end{eqnarray*}
$\epsilon_F$ is calculated by inverting the relation
\begin{eqnarray*}
\sum_{\mathbf{k}}n_{1,\mathbf{k}}+n_{2,\mathbf{k}}&=&\frac{m^*}{2\pi\hbar^2}\int_0^{\infty}d\epsilon \frac{1}{\exp{\beta(\epsilon-\epsilon_F)}+1}+\frac{1}{\exp{\beta(\epsilon+E_{12}-\epsilon_F)}+1}.\\
\end{eqnarray*}

Discretizing the electronic and photonic wavevectors on a grid of
respectively $N_{\mathbf{k}}$ and $N_{\mathbf{q}}$ points, we have
a system of $2N_{\mathbf{k}}+N_{\mathbf{q}}$ equations that can be
numerically solved, e.g., with a Newton algorithm.

\end{widetext}


\begin{thebibliography}{}
\bibitem{ISB-Book} {\it Intersubband Transitions in Quantum Wells:
  Physics and Device Applications I}, edited by H. C. Liu and F. Capasso, Semiconductors and Semimetals
Vol. 62 (Academic Press, San Diego, 2000).
\bibitem{Dini_PRL} D. Dini, R. Kohler, A. Tredicucci, G. Biasiol, and L. Sorba,
Phys. Rev. Lett. {\bf 90}, 116401 (2003).
\bibitem{Aji_APL} A. A. Anappara, A. Tredicucci, G. Biasiol, L. Sorba, Appl.
Phys. Lett. {\bf 87}, 051105 (2005).
\bibitem{Aji_2006} A. A. Anappara, A. Tredicucci, F. Beltram, G. Biasiol, L. Sorba,
Appl. Phys. Lett. {\bf 89}, 171109 (2006).
\bibitem{Ciuti_vacuum} C. Ciuti, G. Bastard, I. Carusotto, Phys.
Rev. B {\bf 72}, 115303 (2005).
\bibitem{Ciuti_PRA} C. Ciuti, I. Carusotto, Phys. Rev. A {\bf 74},
  033811 (2006).
\bibitem{Simone_PRL} S. De Liberato, C. Ciuti, I. Carusotto, Phys. Rev. Lett. {\bf 98}, 103602 (2007).
\bibitem{Faist} J. Faist, F. Capasso, D.L. Sivco, C.  Sirtori,
  A. L. Hutchinson, A.Y. Cho, Science {\bf 264}, 5553 (1994).
\bibitem{Terahertz}   R. K\"{o}hler, A. Tredicucci, F. Beltram,
  H.E. Beere, E.H. Linfield, A.G. Davies, D.A. Ritchie,
R.C. Iotti, F. Rossi, Nature {\bf 417}, 156 (2002).
\bibitem{Raffaele} R. Colombelli, K. Srinivasan, M. Troccoli,
  O. Painter, C. F. Gmachl,
D. M. Tennant, A. M. Sergent, D. L. Sivco, A. Y. Cho, F. Capasso,
  Science {\bf 302}, 1374 (2003).
\bibitem{SST} R. Colombelli, C. Ciuti, Y. Chassagneux, C.
Sirtori, Semicond. Sci. Technol. {\bf 20}, 985 (2005).
\bibitem{Luca_APL} L. Sapienza, A. Vasanelli, C. Ciuti, C. Manquest, C. Sirtori, R. Colombelli, and U. Gennser,
Appl. Phys. Lett. {\bf 90}, 201101 (2007).
\bibitem{Luca-unpublished} L. Sapienza, A. Vasanelli, R. Colombelli, C. Ciuti, Y.Chassagneux, C. Manquest, U. Gennser, C. Sirtori, Phys. Rev. Lett. {\bf 100}, 136806 (2008).
\bibitem{Deliberato_exact} S. De Liberato and C. Ciuti, arxiv:0802.4091. 
\bibitem{F1} J. Fricke, Ann. Phys. {\bf 252}, 479 (1996).
\bibitem{F2} J. Fricke, V. Meden, C.Wohler, K. Schonhammer, Ann. Phys. {\bf 253}, 177 (1997).
\bibitem{KK1} M. Kira, W. Hoyer, T. Stroucken, S. W. Koch, Phys. Rev. Lett. {\bf 87}, 176401 (2001).
\bibitem{Luca_thesis} L. Sapienza, (Ph. D. thesis, University of Paris 7, 2007).
\bibitem{Coulomb}     D. E. Nikonov, A. Imamoglu, L. V. Butov, and H. Schmidt, Phys. Rev. Lett. {\bf 79}, 4633 (1997).
\bibitem{pereira}M. F. Pereira, Jr. , Phys. Rev. B {\bf 75}, 195301, (2007).

\end{thebibliography}
\end{document}